\documentclass[amsmath,amssymb,aps,prd,showpacs,nofootinbib]{revtex4}
\usepackage{graphicx}
\usepackage{amsfonts}
\usepackage{amssymb}
\usepackage{amsmath}
\usepackage{bm}
\usepackage{latexsym}
\usepackage{textcomp}
\newcommand{\beq}{\begin{eqnarray}}
\newcommand{\eeq}{\end{eqnarray}}
\newcommand{\be}{\begin{equation}}
\newcommand{\ee}{\end{equation}}
\def\la{\mathrel{\mathpalette\fun <}}
\def\ga{\mathrel{\mathpalette\fun >}}
\def\fun#1#2{\lower3.6pt\vbox{\baselineskip0pt\lineskip.9pt
\ialign{$\mathsurround=0pt#1\hfil ##\hfil$\crcr#2\crcr\sim\crcr}}}

\newcommand{{\SD}}{\rm SD}

\newcommand{\vep}{\bm p}

\begin{document}

\title{Dynamics of the quark-antiquark interaction and the universality
of Regge trajectories}

\author{\firstname{A.M.}~\surname{Badalian}}
\email{badalian@itep.ru}
\affiliation{Institute of Theoretical and Experimental
Physics, Moscow, Russia}

\author{\firstname{B.L.G.}~\surname{Bakker}}
\email{b.l.g.bakker@vu.nl}
\affiliation{Faculty of Science,
Vrije Universiteit, Amsterdam, The Netherlands}

\date{\today}

\begin{abstract}
The dynamical picture of a quark-antiquark interaction in light
mesons, which provides linearity of radial and orbital Regge
trajectories (RT), is studied with the use of the relativistic
string Hamiltonian with flattened confining potential (CP) and taking
into account three negative corrections: the gluon-exchanged, the self-energy,
and the string corrections. Due to the flattening effect the radial slope
$\beta_n$ and the orbital slope $\beta_l$ of the Regge trajectories decrease
by $\sim 30\%$ as compared to those in linear CP, while the string correction
decreases only the orbital slope by the value $\sim 10\%$. The self-energy correction
is very important and has large magnitude, $\sim -300$~MeV for high excitations. It
also provides the linearity of the RT, built for the centroid squared masses, and
gives the small value of the intercept, $\beta_0=0.50(1)~$GeV$^2$, equal
to the squared centroid mass of $\rho(1S)$. If the universal
gluon-exchanged potential without fitting parameters and screening
function, as in heavy quarkonia, is taken, then the radial slope,
$\beta_n=1.15(9)$~GeV$^2~(l\not= 0)$, and the orbital slope,
$\beta_l=1.03(9)$~GeV$^2$, have close values and the RT can be considered
as approximately universal.

\end{abstract}

\maketitle

\section{Introduction}
\label{sec.1}
The spectroscopy of light mesons refers to the field where
non-perturbative QCD dominates and the Regge trajectories (RT),
both orbital and radial,  appear to be the most explicit manifestation
of non-perturbative effects. It is known that the leading RT in the
($M^2,J$)-plane has a linear behavior with the slope $\beta_J({\rm
exp.})=2\pi\sigma=1.13(1)$~GeV$^2$, which corresponds to the value
of the string tension $\sigma=0.180(2)$~GeV$^2$  in the string
models \cite{1,2}, and precisely this $\sigma$ has been  used in
the realistic potential model with linear confining potential (CP)
\cite{3,4}. Also, systematization of radial excitations of light
mesons, suggested in Ref.~\cite{5}, has shown that their squared
masses lie on linear, or approximately linear, radial trajectories
in the ($M^2,n$)-plane ($n=n_r$ is the radial quantum number) and
has the slope,  $ \beta_n=(1.25\pm 0.15)$~GeV$^2$ \cite{5}. Later
in Refs.~\cite{6,7} a smaller slope $\beta_n=(1.143\pm 0.013)$~GeV$^2$
was extracted from the Crystal Barrel data \cite{8}.

It was also observed  that the slopes of the ($M^2,J$)-trajectories
for the masses with spin $S=0$  and $S=1$ differ only  $\sim 10\%$
\cite{9} and it was assumed that within this accuracy a universal RT
can exist in the  $(l,n)$-plane,
\be
 M^2 (n,l)= a (l + n) + c,
\label{eq.1}
\ee
with the  universal slope $a=1.10(2)$~GeV$^2$ and the intercept
$c=0.68$~GeV$^2$. From  Eq.~(\ref{eq.1}) it follows that the masses
of resonances with equal quantum number $N=l+n$ have  to be equal
and this assumption agrees with the experimental masses of  the
vector resonances with $N=2,3,4~(S=1)$ (see Table~\ref{tab.01}).
%
\begin{table}[!htb]
\caption{The PDG masses  of the isovector resonances (in MeV) \cite{10}
and $M(n,l)$ according to Eq.~(\ref{eq.1})}
 \begin{center}
\label{tab.01}
\begin{tabular}{|l|c|c|c| }
\hline
             &       $N=2$          &    $N=3 $             &  $N=4$   \\
\hline
Meson   & $\rho_3(1690)$ & $\rho_3(1990)$  &   $\rho_3(2250)$ \\
Mass     &  1689(2)             &   1982(14)           &   ~2234  \\
Meson   & $a_2(2P)$    &   $a_4(2040)$     &     $\rho_5(2350)$  \\
Mass      &  1705(40)    &    1995(10)           & 2330(35)   \\
\hline
\end{tabular}

\end{center}
\end{table}

However, in another analysis of the experimental data, where the
PDG masses and widths were used, a larger $ \beta_n=(1.35\pm
0.04)$~GeV$^2$ was extracted \cite{11} and later, after re-analysis
of the experimental data, the same authors have obtained a smaller
$\beta_n=1.28(5)$~GeV$^2$ \cite{12} with the conclusion that the
universality of the radial and orbital RTs is not fulfilled at the
level of 2.4 standard deviations. These results, irrespective of
the fact whether slopes of radial and orbital RTs are equal or
not, raise an important theoretical issue, namely, what dynamical
effects are responsible for the values of the slopes, observed in
experiments, and whether a universal RT exists or not. At present new
studies of the RT nature continues \cite{13,14}.

A study of the light meson spectra in relativistic models shows
that at first sight the RT parameters depend on the quark-antiquark
potential $V_0(r)$ used, but, as shown in the relativistic string
model \cite{15,16,17,18,19,20}, some additional corrections to the
meson masses exist. The potential $V_0(r)$ was studied on a fundamental
level in lattice QCD \cite{21,22} and the field correlator method
\cite{23} in the region $r \la 1.2$~fm. It was shown that in this region
$V_0(r)$ is the sum of the linear confining potential (CP) $V_C(r)=\sigma
r$ and the gluon-exchange (GE) term: $V_0(r)=V_C(r) +V_{\rm GE}(r)$.
Precisely such a linear CP with string tension $\sigma=0.18$~GeV$^2$,
fixed by the slope of leading angular-momentum RT, was used in the
relativistic models \cite{3,4}, where a good description of the masses
of low-lying states was obtained.  However, to describe high
excitations of light mesons, which sizes $\geq 1.5$~fm are large,
knowledge of the quark-antiquark potential at large distances is needed,
which is not defined yet on fundamental level, and in  lattice
QCD a flattening of the CP at $r \ga 1.2$~fm is seen with large
uncertainties. This flattening  (screening) effect appears due to the
creation of  light $q\bar q$ holes (loops) in the Wilson loop and
decreases  the surface of the Wilson loop \cite{17}. However, this
effect was described only at a phenomenological level, assuming that
a study of the high excitations can give important information about the
$q\bar q$ interaction at large $r$ \cite{15,16,17}.

In light mesons one can use the universal GE potential, in which
the GE potential,  $V_{\rm GE}=-\frac{4\alpha_{\rm V}(r)}{3 r}$, is now
well defined at small distances, since at present the QCD constant
$\Lambda_{\overline{MS}}$, as well as the vector constant $\Lambda_{\rm
V}$) is known with a good accuracy for the number of  flavours
$n_f=3,4,5$ \cite{24}. In particular, the value of
$\Lambda_{\overline{MS}}(n_f=3)=315(15)$ MeV \cite{24}, or the corresponding
$\Lambda_{\rm V}(n_f=3)=500(20)$~MeV \cite{25}, appears to be larger
than the value used in the past. However, the behaviour of the strong coupling
$\alpha_{\rm V}$ at small momenta and at large distances, as in the
case of the CP, is still not determined \cite{26} and it remains unclear
whether a screening GE effect exists or not. This problem will be
discussed in our paper.

In Refs.~\cite{15,17} it was shown that the main contribution to the
light meson mass comes from the CP and as a first step it is instructive
to consider the light meson spectrum, taking the purely linear CP at all
distances, and after that to take into account the flattening effect
and other corrections. To make the theoretical analysis more clear we
consider only iso-vector light mesons with $l\leq 3$ and
pay special attention to calculations of the centroid masses. For
that we use the relativistic string Hamiltonian (RSH), which describes
the QCD string  with spinless quarks at the ends and $m_q=0$
\cite{18,19,20}, while the spin-dependent interaction is taken as a
perturbation; in this case the instantaneous $q\bar q$ potential
reduces to the linear plus the GE term.

The RSH is rather complicated and has different representations for large
$l$ and small $l$, $l\leq 3$. Its basic term  $H_0$, given by
\be
 H_0= 2(\sqrt{\vep^2}) +V_0(r),~(l\leq 3)
\label{eq.2}
\ee
is well known and widely used. Its eigenvalues (e.v.s) $M_0(nl)$
can be approximated by an analytical expression with great accuracy,
if the light quark mass $m_q=0$. Notice that if in relativistic
potential models the constituent quark mass, $\tilde{m}_q\sim
(150-200)$~MeV, is used, then the parameters of the RTs depend on the
value of the constituent quark mass.

The e.v. $M_0(nl)$ provides a basic contribution to the meson mass
and for the purely linear CP the squared mass $M_0^2(n,l)$ can be
approximated with great accuracy by the expression  \cite{15,16},
\be
 M_0^2(n,l) = \sigma( 8~l + 4\pi~n + 3\pi \xi(nl) ),
\label{eq.3}
\ee
with $\xi(nl)= 1.0$ with the exception of $\xi(1S)\approx \xi(1P)\approx 1.05$.
From the conventional representation of the RT as
\be
 M^2(n,l)  = \beta_l ~l  + \beta_n~ n + \beta_0,
\label{eq.4}
\end{equation}
and using Eq.~(\ref{eq.3}) for the purely linear CP, one obtains the
following slopes and the intercept,
\be
\beta_l=8 \sigma, ~\beta_n =4\pi\sigma, ~ \beta_0= 3\pi\sigma.
\label{eq.5}
\ee
Now the following problem arises:  if the conventional value of
$\sigma=0.180$~GeV$^2$ is taken, then all parameters of the RT (\ref{eq.5})
are significantly larger that those extracted from the experimental
data \cite{5,6,11,12}. Namely, the orbital slope $\beta_l=1.44$~GeV$^2$
is by $21\%$ larger than $\beta_l(\rm exp.)=1.13$~GeV$^2$ of the leading
RT.  The radial slope $\beta_n=2.26$~GeV$^2$ is about two times
larger than the experimental $\beta_n(\rm exp.)\sim 1.2(1)$~GeV$^2$
for the states with $l\not= 0$ \cite{5,6} (and 1.5 times larger
than $\beta_n(l=0)$ for the radial $\rho-$trajectory), while the
intercept $\beta_0=3\pi \sigma=1.696$~GeV$^2$ is  2.5 times larger
than the corresponding one in Eq.~(\ref{eq.1}.) Notice that the value of the intercept
cannot be decreased introducing a negative (fitting) constant to
the potential $V_0$, as it is often done in potential models.
Moreover,  appearance of this constant in the mass (or potential)
violates the linearity of the RT. It is important that in the RSH, used
here, the $q\bar q$ potential does not contain a fitting constant.

Our goal here is to understand what effects are responsible for the strong
decrease of the intercept and the slopes of the RT (\ref{eq.5}), to
establish the interrelation between the parameters of the RTs and the
potential $V_0(r)$, and to show the role of the string and the
self-energy corrections, which are present  in the mass formulas.
In contrast to our previous analysis \cite{15,16} we do not assume
here that a screening of the GE potential $V_{\rm GE}$ takes place at distances 
$r < 1.2$~fm and this assumption agrees with the results of Ref.~\cite{27}, where it was shown 
that the screening effect of the GE potential is not seen
at distances $r < 1.0$~fm. We also consider how the parameters of the
RT's change for strong and weak vector coupling, taken in  $V_{\rm GE}$.

Our analysis is restricted to orbital excitations with $l \leq 3$,
because high orbital excitations with $l > 3$ have to be considered
in another approximation of the RSH, where the string corrections are very
large and cannot be considered as a perturbation \cite{20},
and  the ground state masses are described by the
expression  $M^2(l,n_r=0)=2\pi\sigma\sqrt{l(l+1)}$, in which the
orbital slope of the leading RT $\beta_l(n_r=0)$  agrees with the
experimental number $2\pi\sigma$, if $l\geq 3$.

We pay special attention to the negative correction produced by the self-energy (SE)
term \cite{28},  which magnitude remains large, $\delta_{\rm SE}\sim -300$~MeV,
even for high excitations of light mesons; being  proportional to
$1/M(nl)$, it provides linearity of the RT.

\section{The mass formulas}
\label{sec.2}

Here we present the structure of the mass formula, using the
simplified version of the RSH,  where the spin-dependent potentials,
as well as the self-energy and the string contribution, are considered as a perturbation
and the values of the angular momentum are restricted to $l\leq 3$
\cite{15,16}. This  RSH $H$ with $m_q=0$,
\begin{equation}
 H =\mu +  \frac{\vep^2}{\mu}  + V_0(r),
\label{eq.6}
\end{equation}
is expressed via the variable $\mu$, determined by the extremum
condition, $\frac{\partial  H}{\partial \mu} = 0$.  It gives
$\mu=\sqrt{\vep^2}$, i.e., $\mu$ is the kinetic energy of a quark.
Then the Hamiltonian $H$ reduces to the form $H_0$ Eq.~(\ref{eq.2})
and its e.v.s are defined by the spinless Salpeter equation (SSE),
\begin{equation}
 (2\sqrt{\vep^2}  + V_0(r))\varphi_{nl}(r) = M_0 (nl) \varphi_{nl}(r).
 \label{eq.7}
\end{equation}
The e.v. $M_0(nl)$ is an important part of the centroid mass
$M_{\rm cog}(nl)$ and for instantaneous $q\bar q$ interaction the potential
$V_0(r)$ is taken as the sum of confining and the GE terms,
\be
V_0(r) = V_{\rm C}(r) + V_{\rm GE}(r),
\label{eq.8}
\ee
where the linear CP  $V_{\rm C}(r)$,
\be
 V_{\rm C}(r)= \sigma r,~ \quad\sigma=0.18~{\rm GeV}^2,
\label{eq.9}
\ee
and also a flattened (screened) CP will be used,
\be
 \quad V_{\rm f}(r)=\sigma_{\rm f}(r) r .
\label{eq.10}
\ee
Here the function $f(r)$ will be given in Sec.~\ref{sec.6}. The conventional form of
$V_{\rm GE}(r)$ is
\be
 V_{\rm GE}(r) = - \frac{4\alpha_{\rm V}(r)}{3 r},
\label{eq.11}
\ee
if there is no a screening effect, and the problem of the GE screening
will be discussed in Sec.~\ref{sec.5}. The contributions from the
GE potential to the masses of excited states are not large, $\la 90$~MeV, nevertheless,
the GE correction is very important, decreasing all parameters of the RTs.

The masses can be calculated by two ways: either solving Eq.~(\ref{eq.7})
with the potential $V_0(r)=V_{\rm C}(r) + V_{\rm GE}$, or considering
$V_{\rm GE}(r)$ as a perturbation. It can be shown that for high
excitations  the exact and approximate values of mass coincide within
$\sim 10$~MeV. Then in the RSH the centroid mass $M_{\rm cog}(nl)$  includes the e.v.
$M_0(nl)$ and three negative corrections: the self-energy, the
string, and $\delta_{\rm GE}$,
\be
 M_{\rm cog}(nl) = M_0(nl) + \delta_{\rm GE} + \delta_{\rm str}(nl) +
 \delta_{\rm SE}.
\label{eq.12}
\ee
where the self-energy correction is the largest one and three
corrections together give a large negative contribution, $\sim -(400
-500)$~MeV, while the e.v.s of the ground states ($n=0$) are the
following: $M_0(1S)=1.339$~GeV,~ $M_0(1P)=1.792$~GeV,~$M_0(1D)=2.155$~GeV.
It is worth to underline that the centroid  mass  $M_{\rm cog}(nl)$
does not contain a fitting negative constant $C_0$, usually introduced
in potential models; this constant produces a non-linear term $C_0
M_0(nl)$ in the squared mass and violates the linearity of RT. On the
contrary, in our approach a negative contribution from the self-energy
correction, $\delta_{\rm SE}(nl)$
\begin{equation}
 \delta_{\rm SE}(nl) = - \frac{\eta_f \sigma}{\mu(nl)}, ~~\eta(q\bar q)=0.90,
\label{eq.13}
\end{equation}
is proportional to $1/M_0$ via $\mu(nl)$ (see below) and therefore
a non-linear term does not appear in the RT. In Eq.~(\ref{eq.13}) the
number $\eta_f$ depends on the quark flavour and in light mesons
we take $\eta_{f=q}=0.90$ \cite{28}. The situation is different in
heavy quarkonia, where  the self-energy  term is small and usually
neglected, since e.g. in bottomonium $\eta_b\sim 0.1$, $\mu_b(nl)\sim
5$~GeV, and $\delta_{\rm SE}\sim -3$~MeV. On the contrary, in a light
meson $\delta_{\rm SE}$ has large magnitude, $\sim -(300-400)$~MeV,
because the kinetic energy m.e. is small. It is important that this
correction slightly decreases in higher excitations, but still
remains large.

Another negative correction, the  string correction $\delta_{\rm str}(nl)$, ($l=1,2,3$) \cite{15,16},
given by
\begin{equation}
 \delta_{\rm str}(nl)= - \frac{l(l+1)\sigma \langle r^{-1} \rangle}{8 \mu^2(nl)},
\label{eq.14}
\end{equation}
increases for states with growing $l$ and decreases for larger  $n$,
however, its magnitude  $\sim  (-40,- 80,-110)$~(in MeV) for $l=1,2,3~
(n=0)$ is not large.  Notice that the expression of $\delta_{\rm
str}$, Eq.~(\ref{eq.14}) does not change if a flattened CP is taken,
but in this case the string tension  $\sigma$ has to be replaced
by the averaged m.e. $\langle\sigma_{\rm f}(r)\rangle_{nl}$, which
is different for every state and smaller than $\sigma$.

For high excitations with $l\not= 0$ knowledge of the centroid mass
is very important, since due to large sizes  their fine-structure
splitting are small and $M_{\rm cog}(nl)$ practically coincides
with the masses of the members of the multiplet.  It does not refer
to  low $nS$ ($1P$) states, where the spin-spin (fine-structure)
splitting is not small. In particular, in the $n\,^3S_1$ states the
hyperfine  correction, equal to -$\frac{1}{4}\delta_{\rm hf}(nS)$, with
\be
 \delta_{\rm hf}(nS) = \frac{8}{9} \alpha_{\rm hf} \tau(nS),~{\rm with}
 ~\tau(nS)= \frac{|R_{nS}(0)|^2} {\mu^2(nS)},
\label{eq.15}
\ee
is not small even for the $4\,^3S_1$ resonance. Calculations show
that the ratio  $\tau(nS)$, Eq.~(\ref{eq.15}),  weakly depends on
the parameters of the GE potential, e.g. for the ground $1S$ state
$\tau(1S)=(0.85-1.05)$~GeV is obtained for different types of GE
potentials. This fact  allows to extract  $M_{\rm cog}(1S)$ from
experiment with an accuracy $\sim 10$~MeV (see below).

Notice that knowledge of $M_{\rm cog}(1S)$ is of special importance
since it determines the intercept of  the leading $l-$trajectory
($n=0$),
\be
 M_{\rm cog}^2(l,n=0) = \beta_l ~l + \beta_{\rm cog}, ~(n=n_r=0),
\label{eq.16}
\ee
with the intercept $\beta_{\rm cog}=M_{\rm cog}^2(1S)$, where
$M_{\rm cog}(1S)=M(\rho(1S)) - \frac{1}{4} \delta_{\rm hf}(1S)$.
In Eq.~(\ref{eq.15})  the  hyperfine correction can be determined
with $\sim 10$~MeV accuracy, if the universal hyperfine coupling $\alpha_{\rm
hf}=0.33(1)$, the same as in heavy-light mesons and bottomonium
\cite{29}, and the theoretical number  $\tau(1S)=0.95(10)$~GeV is
used. It gives $\delta_{\rm hf}(1S)=280(25)$~MeV and  $M_{\rm
cog}(1S,{\rm exp.}) = (775 - \frac{1}{4} 280(25))= 705(6)$~MeV, so
that the ``experimental"  intercept,
\be
 \beta_{\rm cog}({\rm exp.}) = (0.705(6))^2 ~{\rm GeV}^2=0.50(1)~{\rm GeV}^2,
\label{eq.17}
\ee
is smaller than the intercept of the leading RT in the $(M^2,J)$-plane,
defined by the mass of $\rho(1\,^3S_1)$: $\beta_0({\rm exp.})=
M^2(\rho(1S,{\rm exp.}))=0.60$~GeV$^2$.

Notice that for radial excitations the difference between the squared masses,
$b_n^2=M_{\rm cog}^2(n+1,l) - M_{\rm cog}^2(n,l)$, of neighbouring
states can depend  on the radial quantum number $n$. If for all
states with a given $l$ the numbers $b_n^2=b$ are equal, then the
radial RT reduces to the radial RT, introduced in Ref.~\cite{5}:
\begin{equation}
  M(n,l)^2= M_{\rm g}^2 +  b\,n,~ (l~{\rm fixed}),
\label{eq.18}
\end{equation}
where  $M_{\rm g}(n=0,l)$ is the mass of the ground state.

\section{Linear confining potential}
\label{sec.3}

The simplest way to show the structure  of the RTs is to determine
the light meson spectrum in a purely linear CP and
consider  other interactions as a perturbation; in this case the mass $M_{\rm cog}$
is defined by analytical expressions. Notice that the
linear CP plays a special role in  string theory
as well as in  the AdS approach \cite{30}. In a linear potential
the mass formula is simplified owing to the relations,
\be
 M_0(nl) = 4\mu_0(nl), ~~\sigma \langle r \rangle_{nl} =2 \mu_0(nl) = 1/2 M_0(nl).
\label{eq.19}
\ee
In Table~\ref{tab.02} we give the sizes $\langle \sqrt{r^2}\rangle_{nl}$, the
m.e.s $\langle r^{-1}\rangle_{nl}$, and the e.v.s $M_0(nl)$, solving
Eq. (\ref{eq.7}) with the linear  potential $V_{\rm C}(r)$ with
$\sigma=0.180$~GeV$^2$.

\begin{table}[!htb]
\caption{The eigenvalues $M_0(nl)$~(in GeV), the m.e.s
$\langle \sqrt{r^2} \rangle_{nl}$~(in fm),  $\langle r^{-1} \rangle_{nl}$ (in GeV) of
Eq.~(\ref{eq.7}) with the linear potential $V_c(r)=\sigma r$,
$\sigma=0.18$~GeV$^2$\label{tab.02}}

\begin{center}
\begin{tabular}{|c|c|c|c|}
\hline
 State   & $M_0(nl)$ & $\langle \sqrt{r^2} \rangle_{nl}$ & $\langle r^{-1} \rangle_{nl}$\\
  $(n+1)L$  & &  &\\
\hline
1S            & 1.339       & 0.82     &   0.364 \\
2S            & 1.998      &1.26       &   0.330 \\
3S          &  2.498       &1.58        &  0.296 \\
4S            &  2.915     &  1.85      &  0.273 \\
\hline
1P           & 1.792        & 1.06     &   0.236 \\
2P           &  2.315       & 1.43      &  0.226  \\
3P          & 2.750      &  1.72      &    0.214  \\
4P          & 3.129       & 1.97       & 0.204  \\
\hline
1D            &   2.155      & 1.24     & 0.187 \\
2D            &  2.601         & 1.57  & 0.182  \\
3D             & 2.990       &  1.84  &  0.176   \\
4D            &   3.337      & 2.08    &  0.170  \\
\hline
1F             & 2.465       & 1.41    & 0.159 \\
2F              & 2.861      & 1.71     & 0.157  \\
3F              &  3.215   & 1.96       & 0.153  \\
4F             &   3.538     &  2.18    & 0.149\\
\hline\end{tabular}
\end{center}
\end{table}
Knowing the m.e.s $\langle r^{-1} \rangle_{nl}$ and the e.v.s $M_0(nl)$, we have observed 
that in a purely  linear CP the m.e.s  $\langle r^{-1} \rangle_{nl}$ can be approximated with
an accuracy better than $2\%$ as
\be
\langle r^{-1} \rangle_{nl} = M_0(nl) A(nl),
~A(nl)= \frac{0.262 (l+2)}{(l+1)(l+n+2)},~( l\not= 0);~
\langle r^{-1} \rangle_n =  M_0 A_0(n),
~ A_0(n)=2\frac{0.271}{n+2}= \frac{0.542}{n+2},~(l=0),
\label{eq.20}
\ee
i.e., they are proportional to $M_0(nl)$. Then, with the use of the
relations (\ref{eq.19}) and (\ref{eq.20}) all corrections to $M_{\rm
cog}(nl)$ are given by analytical expressions.

For further analysis we rewrite the expression of $M_0(nl)^2$
(\ref{eq.3}) with $\sigma=0.180$~GeV$^2$,
\be
 M_0^2(nl)({\rm in~GeV}^2) =  (1.440\, l + 2.262\, n + 1.696\,\xi(nl)),
\label{eq.21}
\ee
where the numbers  $\xi(nl)=1.0$ with an accuracy better $2\%$  for
all states,  with the exception of $\xi(1S)=1.057$  and
$\xi(1P)=1.045$. Note that in $M_0^2(nl)$ (\ref{eq.21}) the slopes
$\beta_l=1.44$~GeV$^2$,~$\beta_n=2.26$~GeV$^2$, and  the intercept
$\beta_{\rm cog}=1.70\xi$~GeV$^2$ are significantly larger than
those, extracted from experimental data \cite{5,6,11,12}, while due
to the GE, the string, and the SE corrections the masses $M_{\rm
cog}(nl)$ and the parameters of the RT decrease.

Then with the use of Eqs.~(\ref{eq.19}) and (\ref{eq.20}) the
orbital slope decreases owing to the string correction Eq.~(\ref{eq.14}),
\be
\beta_l= \sigma \left( 8 -1.048 \frac{l(l+2)}{l+2+n} \right) ~{\rm for}~ M=M_0 + \delta_{str},
\label{eq.22}
\ee
and in the general case it depends on the quantum number $l$:
$\beta_l(n=0)=1.251$~GeV$^2$ for $l=1$ and  $\beta_l(n=0)=1.067$
GeV$^2$ for $l=2$; for the radial RT with $n=1$  the slope
$\beta_l(n=1)=1.299$~GeV$^2$ for $l=1$ and $\beta_l(n=1)=1.138$~GeV$^2$
for $l=2$; for the daughter RT with $n=2$
$\beta_l(n=2)=1.327$~GeV$^2~(l=1)$ and $\beta_l(n=2)=1.188$~GeV$^2$
for $l=2$. Thus with the string correction taken into account the
orbital slope remains large and $l$-dependent, i.e., the  RT's can
be considered as approximately linear.

\subsection{The GE correction to the centroid mass}

Here we take the GE potential as a perturbation and later show that
exact solutions of the SSE with $V_(0)=V_{\rm C}(r)+ V_{\rm GE}$ give a
contribution to the mass, which coincides with the GE correction
with high accuracy (see section \ref{sec.6}). Using Eq.~(\ref{eq.20}) the GE
correction (\ref{eq.11}) can be rewritten as ($e_{\rm eff} = \frac{4}{3}\alpha_{\rm eff.} $)
\be
\delta_{\rm GE}= - \frac{4}{3}\alpha_{\rm eff.}(nl)\langle r^{-1}\rangle_{nl} =
 - e_{\rm eff} M_0(nl)\, A(nl), ~
\label{eq.23}
\ee
where in general  the effective coupling,
$\alpha_{\rm eff}(nl)=\langle \alpha_{\rm V}(r) \rangle $ depends on the
quantum numbers $n$ and $l$.  However, in high excitations this dependence
becomes weak because of their large sizes, $\ga 1.4$~fm, and the
m.e.s  $\alpha_{\rm eff}(nl)$ are practically equal for all states,
with the exception of the $1S,2S$ and $1P$ ground states, for which the
asymptotic freedom (AF) behavior of the coupling is important (see
below). For other states, the values of $\alpha_{\rm eff}(nl)$ appear
to be only  $\sim 3\%$ smaller than the asymptotic coupling
$\alpha_{\rm asym}$. Therefore, for high excitations one can put
$\alpha_{\rm eff}(nl)=\alpha_{\rm asym}$. A typical $\alpha_{\rm
asym}$, used in relativistic models, lies in the range, 0.55-0.63
\cite{3,4,16} . This value was also derived on a fundamental level
\cite{23,25,26}, where the uncertainty  depends on the values of the
vector QCD constant $\Lambda_{\rm V}(n_f=3)$ and the infrared (IR)
regulator taken (see  Section~\ref{sec.5}). With the Coulomb constant
$e_{\rm asym}=\frac{4}{3}\alpha_{\rm asym} \cong 0.72(4)$ and using
the factor $A(nl)$ (\ref{eq.20}), one can see that
\be
 \delta_{\rm GE}(nl) = - e_{\rm asym}  M_0(nl) A(nl), ~(l\not= 0,~n\geq 1),
\label{eq.24}
\end{equation}
is proportional to  the e.v. $M_0(nl)$. It means that the GE
correction gives a negative contribution to all parameters of the RT:
the slopes $\beta_l,~\beta_n$, and the intercept. Then  the mass
$M_{\rm GE}(nl)$ with the GE correction taken into account is
\be
M_{\rm GE}(nl) =M_0(nl)\, Z(nl), ~{\rm with}~Z(nl)=( 1 - e_{\rm asym}  A(nl)),
~(l\not= 0,~n\geq 1).
\label{eq.25}
\ee
but for the $nS$ states
\be
M_{\rm GE}((n+1)S) = M_0((n+1)S)\,Z_0(n), ~ Z_0(n)=(1  - e_0(n) A_0(n)),
\label{eq.26}
\ee
where $e_0(n)\not= e_{\rm asym}$ and the quantities $A_0(n)$ are larger than $A(nl)$
with $l\not= 0$. From Eq.~(\ref{eq.25}) one can see that the
parameters of the RTs can depend on the quantum numbers through the factor
$A(nl)$, but in high excitations this dependence is weak because
the term $e_{\rm asym}\, A(nl)$ is small even for a strong GE potential.
We choose the Coulomb constant, $e_{\rm asym}=0.76$ (or $\alpha_{\rm
asym} =0.57$) (see below) and define  the average
$\langle A(l, {\rm fixed}~n) \rangle =\frac{1}{2} (A(l=1,n)+ A(l=2,n))$.
Then for $n=0$ one finds $e_{\rm asym}\langle A(l,n=0) \rangle = 0.084(16),
~Z(l,n=0)=0.916(16)$ and $Z^2(l,n=0)=0.839(30)$. For $n=1$ with
$e_{\rm  asym}\langle A(l,n=1) \rangle=0.064(11)$,  the factor
$Z(n=1)=0.936(11),~Z^2(n=1)=0.876(20)$ is larger; for
the daughter RT with $n=2$, $\langle A(n=2) \rangle = 0.052(8)$ , $Z^2(n=2)=0.899(14)$
are obtained. We can conclude that in the  linear CP due to the factor $Z^2(nl)$,
defined by the GE correction, the orbital slope decreases  by $\sim (10-16)\%$, but
still remains large, $\beta_l\sim (1.24-1.30)$~GeV$^2$.
Also the intercept decreases, although its value,  $3\pi \sigma\,
Z^2(n)\sim (1.43-1.53))$~GeV$^2$, is kept large for all RTs.

The GE corrections give a contribution to the kinetic energy m.e.s,
denoted as $\mu_{\rm GE}$, see Eq.~(\ref{eq.27}), which are given
in Table \ref{tab.03}  together with the SE and  the string corrections, and $M_{\rm
cog}(nl)$. From this table one can see that for the ground states their
masses agree with experiment, while for the $2S,2P,2D$ states and
higher excitations  the masses $M_{\rm cog}(nl)$ are larger by
$(100-200)$~MeV than experimental values and the only way to decrease these masses
is to take into account a flattened, or screened, CP. Note
that without the SE and the string corrections the masses $M_{\rm
GE}(nl)=M_0 +\delta_{\rm GE}$ are larger by $\sim (300-400)$ MeV
than the experimental masses $M_{\rm cog}(\rm exp.)$.

There exists another effect, produced by the GE potential, which
increases the quark kinetic energy and for the states with $l\not=0$
this m.e.  can be approximated (with an accuracy better than $5\%$) by
\be
 \mu_{\rm GE}(nl) = \mu_0(nl) + \frac{1}{4} \delta_{\rm GE}(nl)=
  \frac{M_0(nl)}{4} (1  + e_{\rm eff}(nl)\, A(nl)).
\label{eq.27}
\ee
The kinetic energy $\mu_{\rm GE}$ is larger than $\mu_0(nl)$  and
has to be taken into account in the self-energy and the string
corrections, which decrease due to this effect. In some cases,
instead of the approximation (\ref{eq.27}), one can use another
approximation for  $\mu_{\rm GE}$,
\begin{eqnarray}
 \mu_{\rm GE}(nl) & = & 1.11\,\mu_0(nl)= 0.275\, M_0(nl),
 \nonumber \\
 \mu_{\rm GE}(n=l=0) & = &1.21\, \mu_0(l=n=0) =0.3025\, M_0(l=n=0).
\label{eq.28}
\end{eqnarray}

\subsection{The string correction}

With the use of the modified kinetic energy $\mu_{\rm GE}$ (\ref{eq.28})
the string correction, proportional to $\langle r^{-1} \rangle_{nl}=A(nl) M_0(nl)$
(\ref{eq.13}), can be written as
\be
\delta_{\rm str}(nl) =  - l (l+1) \frac{\sigma \langle r^{-1} \rangle}{8\mu_{\rm GE}^2}
 = -l (l+1)\, 1.623\, \sigma\, A(nl)\, M_0^{-1},
\label{eq.29}
\ee
i.e., it is proportional to $l$ and contributes only
to the orbital slope  $\beta_l$, decreasing its value. Since the
string correction is not large ($\delta_{\rm str}\sim -45$~MeV for
the $1P$ state,  $\sim -80$~MeV, ~$\sim -105$~MeV, respectively, for
the $1D$ and $1F$ ground states, and smaller for radial excitations (see
Table \ref{tab.03}), it decreases the orbital slope only by $(5-10)\%$.
Nevertheless, taking into account the string correction improves the agreement
of the theoretical $\beta_l$ with the experimental value $\beta_l(\rm exp.)=1.13$~GeV$^2$
\cite{11}.

\subsection{The self-energy correction}

The SE correction (\ref{eq.13}) is of special importance in light
mesons and with the modified kinetic energy Eq.~(\ref{eq.27}) can be
rewritten as
\be
\delta_{\rm SE}(nl) = -\frac{0.9\, \sigma}{\mu_{\rm GE}} = - \frac{3.243\,\sigma}{M_0(nl)},
\label{eq.30}
\ee
being  proportional to $M_0^{-1}(nl)$. Therefore, $\delta_{\rm SE}$
produces a negative constant in the squared mass and strongly
decreases the intercept, but does not change the radial and orbital
slopes.

In Table \ref{tab.03} the centroid mass $M_{\rm cog}(nl)$ (\ref{eq.10}), the
corrections $\delta_{\rm GE},~\delta_{\rm str}, ~\delta_{\rm SE}$,
defined by the Eqs.~(\ref{eq.24}), (\ref{eq.29}), and (\ref{eq.30}),
are given together with the averaged kinetic energy $\mu_{\rm GE}$
(\ref{eq.27}) and the experimental values of $M_{\rm cog}(nL)$, which
are known, if the experimental masses of all members of a multiplet are
measured. In the cases where in the PDG \cite{10} only the mass of the
highest state with $J=l+1$ is given, then an inequality $M_{\rm
cog}(nl,{\rm exp.}) < M(J=l+1,{\rm exp.})$ takes place. For illustration we
have chosen the vector coupling equal to a constant, $\alpha_{\rm V}=0.482$,
or $e=0.643$, and neglected the asymptotic freedom (AF) effect.

\begin{table}[!htb]
\caption{The centroid mass $M_{\rm cog}(nl)$ (\ref{eq.12}), the
kinetic energy $\mu_{\rm GE}(nl)$, the GE correction $\delta_{\rm
GE}$ with $e=0.643$ (in GeV), and the corrections $\delta_{\rm str},
~\delta_{\rm SE}$ (in GeV) in the purely linear confining potential \label{tab.03}}
\begin{center}

\begin{tabular}{|c|c|c|c|c|c|r|}
\hline
State&$\delta_{\rm GE}$&$\mu_{\rm GE}$& $\delta_{\rm SE}$&
$\delta_{\rm str}$ & $M_{\rm cog}(nl)$&$M_{\rm cog},{\rm
exp.}$~\cite{10}\\
$(n+1)L$& & & &  & &\\
\hline
1S            & -0.234         &  0.405        & - 0.400      &   0      & 0.705    & 0.705(6)\\
2S            & -0.212         &  0.553      &   -0.293        & 0       & 1.493   &1.424(25)\\
3S            & -0.182         &  0.672      &   -0.241       &  0       & 2.067    & 1.875(5)\\
4S            &-0.168          &  0.773      &  -0.210         & 0         &  2.529  & absent \\
\hline
1P            &-0.152          & 0.486       &   -0.333       & -0.044     & 1.263   & $< 1.318$\\
2P            &-0.146          &  0.615         & -0.263       & -0.027    & 1.879   & $<1.732(9)$\\
3P            &-0.138          & 0.722       &   -0.224        &  -0.018   & 2.369   & absent \\
\hline
1D            & -0.120          & 0.569    &  -0.285           & -0.076   & 1.674    & $\approx 1.69$ \\
2D            & -0.117          & 0.679  & -0.238              & -0.052    & 2.194    & $\approx 1.990$\\
3D    &          -0.113         & 0.776   & -0.209           & -0.038      & 2.630   & absent \\
\hline
1F             & -0.102    &      0.642   &   -0.252         & -0.102     &  2.009    & $\sim 1.995 (10)$\\
2F             & - 0.101        & 0.741   &  -0.219           & -0.076    &  2.465   & absent \\
\hline
\end{tabular}
\end{center}
\end{table}

In a more realistic case one can take $\alpha_{\rm eff}(nl)= \alpha(\rm asym.)$, i.e.,
$e_{\rm eff}(nl)=e_{\rm asym}$, for all states (with exception of the states
$1S$, $2S$ and $1P$); then  $Z(nl)=1 - e_{\rm asym}\, A(nl))$ and the expression of the
centroid mass $M_{\rm cog}(nl)$ is simplified to
\be
 M_{\rm cog}(nl) = M_0(nl) Z(nl)  -  1.623\, l(l+1)\sigma \frac{A(nl)}{M_0 } - \frac{3.24\,
 \sigma}{M_0},~ (l\not= 0,~n\geq 1).
\label{eq.31}
\ee
Than the squared mass has a clear structure,
\be
 M_{\rm cog}^2(nl) = M_0^2 Z^2(nl) - 3.246\, \sigma l (l+1)\,A(nl)\, Z(nl)  - 6.48\, \sigma Z(nl) +
 \delta_{\rm SE}^2 +{\rm  small~  terms}.
\label{eq.32}
\ee
From  Eq.~(\ref{eq.32}) several conclusions can be drawn.
One can see that the corresponding RT is non-linear through the terms
$A(nl)$  and $Z(nl)$, however, taking the averaged $\bar A=\langle A(n)\rangle$ for a
given $l$, the  radial RT's can be considered as approximately
linear.

In the radial slope the GE correction ($l$ is fixed) is
defined by $Z(l)^2$,
\be
\beta_n({\rm fixed}~ l)= 4\pi \sigma Z^2(l) = 4\pi\sigma (1- e_{\rm eff.} \bar A(l))^2,
\label{eq.33}
\ee
and the value $\beta_n(l)=(1.96-2.06)$~GeV$^2$ remains large for any
$l$, being $\sim 70\%$ larger than the experimental radial slope,
$\beta_n\sim 1.2(1)$~GeV$^2$~($l\not= 0$) \cite{5,6,7}, even if the
strong GE potential is used. Just owing to the large radial slope the
large masses  $M_{\rm cog}(nl)$, given in Table \ref{tab.03}, are obtained.

In  this table we give $M_{\rm cog}(nl)$, calculated with $e_{\rm
eff}=0.643$, which is smaller than $e_{\rm asym}$, and  in this
case $M_{\rm cog}(2S)$  is  $\sim 70$~MeV larger than
$M(\rho(2S)$;  $M_{\rm cog}(2P)$ is larger by $\sim 150$~MeV than
$M(a(1320)$, and $M_{\rm cog}(2D)$ is larger than $\rho_3(1990)$
by $\sim 200$~MeV. These masses would be only  $\sim 30$~MeV smaller,
if  the larger $e_{\rm eff}=e_{\rm asym}=0.76$ was used.

The orbital slope decreases  owing to both the GE and the string
corrections and with $Z(nl)=1 -e_{\rm asym} A(nl)$,
\be
 \beta_l=\sigma \left(8\, Z^2(nl) - 0.851\, \frac{l+2}{l+2+n}Z(nl) \right),
\label{eq.34}
\ee
where for large $l$ and $n=0,1$ the contribution from $\delta_{\rm str}^2$,
which was neglected in Eq.~(\ref{eq.34}), can be not small.

For the leading Regge trajectory (LRT) with $n=0$ the orbital slope ( $\bar
A(l)=0.084(16)),~e_{\rm asym}=0.76$) $\beta_l(n=0)=5.933\sigma=1.07(3)$~GeV$^2$ is
in  good agreement with the experimental value $\beta_l(\rm exp.)=1.13(1)$~GeV$^2$
\cite{7,11}. However, in a daughter RT, e.g. with $n=2$ ($\bar
A(l)=0.052(21),~ e_{\rm asym}\bar A(l)=0.044(16)$) the orbital slope
$\beta_l(n=2)=6.77\sigma=1.22$~GeV$^2$ is by $14\%$ larger
than $\beta_l(n=0)$ and this RT is not parallel to the LRT.

From Eq.~(\ref{eq.32}) one can see that the contributions to the
intercept come from the GE and self-energy corrections. For the LRT
the intercept $\beta_{\rm cog}(n=0)=0.50(1)$~GeV$^2$ was
already determined from the experimental value of $M(\rho(1S)$, while
in the orbital RT with $(n=1)$ ($\bar A(n=1)=0.058(14),~~eA=0.044(11),~Z(n=1)=0.956$)
a cancellation of two terms occurs,
\be
 \beta_{\rm cog}=  \sigma(3\pi 0.956^2 -6.195) = 2.42(18)\sigma = 0.44(3)~{\rm GeV}^2,
\label{eq.35}
\ee
and the calculated intercept agrees with the experimental
intercept, $\beta_{\rm cog}(\rm exp.)=0.50(1)$~GeV$^2$, within the
accuracy of the calculations.

Thus we conclude that in the purely linear CP with all corrections taken into account and
large Coulomb constant, $e\sim (0.64-0.76)$, the masses of
the ground states agree with experiment, while the masses of first excitations exceed
the experimental values by $\sim (100-150)$~MeV.

\section{The leading Regge trajectory}
\label{sec.4}

The leading RT describes the ground states with $S=1$, where
the $1S$, $1P$, and $1D$ states have relatively small sizes (see Table \ref{tab.02}), so
for them the use of the linear CP can be  justified.  In the $(J,M^2)$-plane ($J=l+1$)
the LRT can be written as $M^2(J,n=0)=(1.13\, J - 0.53)$~GeV$^2$,
or in the $(l,M^2)$-plane it can be rewritten similar
to that for $M_{\rm cog}^2(nl)$ (\ref{eq.16}),
\be
 M^2(J=S+1,n=0) = \beta_l\, l + \beta_{0J},~{\rm with}~ \beta_{J0}= M^2(\rho(1S))=0.60~
 {\rm GeV}^2,
\label{eq.36}
\ee
where the intercept  $\beta_{0J}$  is
larger than the intercept $\beta_{\rm cog}=M_{\rm cog}(1S)^2= 0.705(6)^2$~GeV$^2$ = 0.50(1) GeV$^2$
~(\ref{eq.17}).

To determine the intercept of  the LRT it is not sufficient to take into
account the self-energy correction,  otherwise in a purely linear
CP (with  $M_0(1S)=1.339$~GeV and $\mu_0(1S)=\frac{M_0}{4}=0.335$~MeV)
one would obtains the mass  $M_{\rm cog}(1S,{\rm lin}) = M_0(1S)
-\frac{3.6\,\sigma}{M_0(1S)}=0.826$~GeV ($\sigma=0.180$~GeV$^2$),
which is even larger than the experimental mass of the $\rho(1S)$ meson.
As was shown in last section, owing to the GE potential, the kinetic energy increases from
the value $\mu_0(1S)=0.335$~GeV to $\mu_{\rm GE}(1S)= (0.395$ \textdiv $0.415)$~GeV=
0.405(10)~GeV, where  the uncertainty depends on the uncertainty in the
QCD vector constant $\Lambda_{\rm V}(n_f=3)$ taken (see Table \ref{tab.04} and
Sec. \ref{sec.6}), and the value  of $\mu_{\rm GE} =0.405(10)$~GeV is
obtained from the exact solutions of the SSE (\ref{eq.7}).

With $\mu_{\rm GE}(1S)= 0.405(10)$~GeV the self-energy correction,
$\delta_{\rm SE}(1S)= - 0.400(10)$~GeV, decreases, being  $\sim 100$~MeV
smaller than that for $ \mu_0(1S)$.

In the LRT the effective constants  $\alpha_{\rm eff.}(1S) = \langle
\alpha_{\rm V}(r) \rangle_{1S}$ are not equal for all states, since
the AF effect decreases $\alpha_{\rm eff.}$ for the $1S$ and $1P$ states by
$\sim (10-15)\%$, while for the states with $l=2,3$ their couplings
are practically equal to the asymptotic coupling $\alpha_{\rm asym}$.

It is of interest to notice that the coupling $\alpha_{\rm eff.}(1S)$ can be extracted from
experiment, if one uses the ''experimental" value of the centroid
mass, $M_{\rm cog}(1S,{\rm exp.})=0.705(6)$~GeV (\ref{eq.17}).
Taking $\mu_{\rm GE}=0.405(10)$~GeV,  $\delta_{\rm SE}(1S)=0.400$~GeV,
and the mass $M_{\rm cog}(1S)$ given by
\be
M_{\rm cog}(1S) = M_0(1S) + \delta_{\rm SE}(1S) + \delta_{\rm GE}(1S) =
(1.339 - 0.400(10) - e_0 (1S)\langle r^{-1} \rangle_{1S}) ~{\rm GeV} = 0.705(6),
\label{eq.37}
\ee
one determines the effective constant $e_0(1S)=0.643(41)$ (here $\langle
r^{-1} \rangle_{1S} =0.364$~GeV), or the effective coupling,
$\alpha_{\rm eff.}(1S)=0.482(31)$,  with a theoretical error $\sim
6\%$. Note  that the lower limit of  $alpha_{\rm eff.}(1S)=0.45$, which
appears to be significantly larger than $\alpha_{\rm V}=0.30$, used in our
paper before \cite{15}. At the same time the upper limit of this coupling,
equal to 0.51, is smaller than $\alpha_{\rm asym}=0.57(3),~e_{l\,{\rm
asym}}=0.76$, used in high excitations, confirming the influence of the
AF effect.

For the ground $1S$ state from Eq.~ (\ref{eq.20}) one has $A_0(n=0)=0.272$
and with  the fitted value $e_0(1S)=0.643(41)$~($\delta_{\rm GE}(1S)=-0.234(15)$~GeV),
one obtains the factor $e_0(1S)\,A_0(n=0)=0.175(11)$, or
\be
 Z_0(1S)= 1 - e_0 A_0(n=0)=0.825(16),
\label{eq.38}
\ee
i.e., the factor $Z_0^2(1S)$ decreases the squared mass (\ref{eq.37}) by
$\sim 32\%$ and provides the correct value of  the intercept, $\beta_{\rm
cog}(1S)=\beta_{\rm cog}(\rm exp.)= 0.50(1)$~GeV$^2$.
In the ground states with $l\not= 0$ the GE correction (\ref{eq.24}), proportional
to $M_0(l,n=0)$,
\be
 \delta_{\rm GE}(1l)= - 0.262\, e_l(n=0) \frac{M_0(1l)}{l+1}, ~(l\not= 0,~n=0),
\label{eq.39}
\ee
in general  contains different values of $e_l(n=0)$ and $Z(l,n=0)$
\be
Z(l,n=0) = 1 - 0.262\,\frac{e_l(n=0)}{l+1},  ~~(l\not= 0).
\label{eq.40}
\ee
Then the centroid mass can be rewritten as,
\be
 M_{\rm cog}(1l) = M_0(1l) Z(l,0) - \frac{3.243\,\sigma}{M_0(1l)} -
  \sigma l \frac{0.427}{M_0},~(l\not= 0),
\label{eq.41}
\ee
if the approximate relation  $\mu_{\rm GE}=1.11\mu_0$, following from the Eqs.~(\ref{eq.27}) 
and (\ref{eq.28}),  is used, then the squared mass $M_{\rm cog}^2(1l)$ is
\be
 M_{\rm cog}^2(1l)= M_0^2 Z^2(l,0) - 0.851\,\sigma l Z(l,0) - 6.486\, \sigma Z(l,0) +
  \delta_{\rm SE}^2 +~{\rm small ~terms}, 
\label{eq.42}
\ee
Here, in the orbital slope the constants $e_l(n=0)$ are different for the $1P$
and the ground states with $l\geq 2$, for which the asymptotic value,
$e_l(l,n=0)=e_{l\,{\rm asym}}$ can be used, while due to the AF effect the
coupling $\alpha_{\rm eff.}(1P)$ has a value close to that for the $1S$ state and
here we take $e_l(,l=1,n=0)=e_0=0.643(41)$ and $e_l(l\geq 2,n=0)=e_{\rm asym}=0.76$.
Then with $A(l=1,n=0)=0.131,~A(l=2,n=0)=0.0873$ and the average,
$\langle e_l A(l,n=0) \rangle=0.075(5)$, and $Z(n=0)=0.925(5)$ from the
Eq.~(\ref{eq.42}) the orbital slope is
\be
 \beta_l(n=0) = \sigma ((8 Z^2(n=0)  -0.851 Z(n=0)) = 6.06(7)\sigma
 =1.09(1)~{\rm GeV}^2,
\label{eq.43}
\ee
which agrees almost precisely  with experimental slope, $\beta_l=1.13(1)$
\cite{7}.

Also with the chosen constants $e_l (1P)=0.643(41)$ and  $Z(1P)=0.916(5)$
the intercept, $\beta_{\rm cog}(1P)= \sigma( 3\pi\, Z(n=0)^2 -
6.484\,Z(n=0)) + \delta_{\rm SE}^2 = 2.58(5)\sigma = 0.46(1)$~GeV$^2$, is
obtained in good agreement with the experimental number, $\beta_{\rm cog}=0.50(1)$, where
a contribution from the squared correction $\delta_{\rm SE}^2=0.11$~GeV$^2$ is $\sim 25\%$.

In Table \ref{tab.03}  for simplicity we give the centroid masses with equal
Coulomb constant, $e_l=e_0 =0.643$, and the masses $M_{\rm cog}(1P)=1263$~MeV,
~ $M_{\rm cog}(1D)=1674$~MeV,~ $M_{\rm cog}(1F)=2009$~MeV (without fine-structure splitting)
turn out to be in good agreement  with the experimental masses of $a_2(1320),~\rho_3(1690)$,
and $a_4(2040)$ (its mass, $M({\rm exp.})=1995^{+10} _{-8}$~MeV)
\cite{10}). This agreement indicates that in the ground states
with $l\geq 2$ the fine-structure splittings are not large. Notice
that the magnitudes of the string corrections, which are equal to
$-48$\,MeV, $-81$\,MeV, and $- 142$\,MeV, for $l=1,2,3$, respectively,
grow for increasing  $l$.

In conclusion in Table~\ref{tab.04} we give the parameters of the RT's for different
types of the potential $V_0(r)$: for the purely linear CP, the linear
CP + weak $V_{\rm GE}$, and for the linear CP+strong $V_{\rm GE}$.

\begin{table}[!htb]
\caption{The averaged values of the orbital and the radial slopes, and
the intercept (in GeV$^2$) of the Regge trajectories for the linear
CP $V_C(r)~(\sigma=0.18$~GeV$^2$) and different gluon-exchange terms
$V_{\rm GE}$ \label{tab.04}}
\begin{center}
\begin{tabular}{|c|c|c|c|}
\hline
   Potential            & linear CP      &linear CP+ weak $V_{\rm GE}$  & linear CP+ strong $V_{\rm  GE}$\\
  Corrections                 &  0         &  $\delta_{\rm SE}\not=0$,  &~ $\delta_{\rm SE}\not=0$   \\
 \hline
$\alpha(\rm eff.)$             & 0            &   0.30                &   0. 57  \\
$\langle\beta_l(n=2)\rangle$               &  1.440       &   1.225(1)             &   1.13(2)\\
$\langle\beta_l(n=1)\rangle$               &   1.440      &   1.17(4)              & 1.13(3)\\
$\langle\beta_l(n=0)\rangle$, $l\not= 0$    &   1.440       &  1.12(3)             & 1.09(2)  \\
$\langle\beta_n\rangle~ (l\not= 0)$        &   2.262       &  2.14(2)              &  1.97 (5) \\
$\langle\beta_{\rm cog}\rangle,~(n=2)$     &  1.696       & 0.47(1)                &  0.46(2) \\
\hline
\end{tabular}
\end{center}
\end{table}

Thus, our analysis of the RTs, when the CP is linear at all distances
and the SE, the string and the GE contributions are taken as a
perturbations, has allowed to get analytical expressions for the
masses and the parameters of the RTs, which have  several characteristic
features:

\begin{enumerate}
\item The radial slope, $\beta_n\cong 2.0$~GeV$^2$ ($l=1,2,3$),
remains larger than $\beta_n(\rm exp.)=1.2(1)$~GeV$^2$ by $\sim
60\%$, irrespective to the strength of the GE potential used, and the
cases with $\alpha_{\rm eff.}=0.30$ and $\alpha_{\rm eff.} =0.57$
were compared.

\item On the contrary, the orbital slope of the LRT
$\beta_l(n=0)=1.09(1)$~GeV$^2$ agrees with the experimental value, if
the strong vector coupling $\alpha_{\rm V}\sim 0.53(4)$ is taken.
This choice of the coupling is preferable, since for small coupling,
$\alpha_{\rm V}=0.30$,  the orbital slope $\beta_l(n=0)\sim 1.20(2)$~GeV$^2$
and the mass of $\rho(1S)$ is larger than in experiment.

\item In the linear CP the orbital slope of the daughter RTs ($n\geq
1$) is  by $(10-15)\%$ larger than $\beta_l(n=0)\approx \beta_l(\rm
exp)=1.13(1)$~GeV$^2$, even if the strong GE potential is used. Precisely
for that reason the $q\bar q$ interaction has to be modified at
large distances.

\item The  largest effect from  the GE potential refers to the
masses of the $nS$ states, which increases the radial slope of the
$\rho(nS)$- trajectory (see section \ref{sec.6}).

\end{enumerate}

\section{The gluon-exchange potential at large distances}
\label{sec.5}

Here we use the conventional $q\bar q$ potential $V_0(r)$ as a
simple sum, Eq.~(\ref{eq.8}). This representation is confirmed by
the Casimir scaling effect, observed in lattice QCD \cite{31} and
derived in the field correlator method \cite{32}. Meanwhile, this
choice as the sum of two terms does not imply that each term, the CP and the GE
potentials, is described by the simple expression as in Eqs. ~(\ref{eq.9}) and
(\ref{eq.11}) at all distances. Moreover, in lattice QCD the linear
behavior of the CP is proved to be valid only in the region $r\la 1.2$~fm,
while for $r > 1.2$~fm, the flattening, or screening, of the CP
is seen, but the details of $V_{\rm GE}$ are not studied yet and the
flattened CP was only  introduced phenomenologically in several models
\cite{15,33,34}.

Also the expression of the GE potential (\ref{eq.11}), taken from
perturbative QCD, in a strict sense is valid only up to the momentum
$q^2\ga 1.5$~GeV$^2$ in momentum space, or down to very small
distances, $r < 0.1$~fm in coordinate space \cite{35,36}.
Therefore, to use $V_{\rm GE}(r)$ in coordinate space in the
whole region, one must first regularize the vector coupling $\alpha_{\rm V}(q^2)$
in momentum space and then regularize
$\alpha_{\rm V}(r)$ by using in Eq.~(\ref{eq.47}) (see below) the regularized
$\alpha_{\rm V}(q^2)$. Notice that
the asymptotic values of $\alpha_{\rm V}$ are equal in  momentum
and  coordinate space \cite{36}. It seemingly supports the idea
that the OGE and confining interactions are not independent.

Here we follow the detailed analysis from Ref.~\cite{27}, which
reveals at least three important effects, which can modify $V_{\rm C}(r)$
and $V_{\rm GE}$ owing to background fields.

\begin{enumerate}
\item The gauge invariance of the gluon exchange in the confining
background requires the propagating gluon to be inside the confining
film (the surface, filled by the background fields), connecting the
$q$ and $\bar q$ trajectories. The resulting area of the film should
obey the Wilson  minimal-area law \cite{27}.

\item The propagating gluon can create  $gg$ (gluon-gluon) loops
in the confining film only in the higher $O(\alpha_s)$ orders, which
introduces a new mass parameter $M_{\rm B}$ \cite{37},  expressed
via the string tension, and its value, $M_{\rm B}^2=2\pi\sigma$,
defined with $10\%$ accuracy,  enters in the evolution equation
together with $q^2$, making the coupling dependent on the variable
$(q^2+M_{\rm B}^2)$ \cite{16,37}.

\item If the confining string is long, it can create light
$q\bar q$ holes in the confining film, thus decreasing the surface
of the Wilson loop (i.e., the film surface). A finite density of this
holes gives rise to the flattening of the confining potential at
large $r$ and due to the flattening effect the masses of high
excitations decrease, since their effective string tension is smaller
for large $r$ than that in the region $r < 1.2$~fm \cite{15,16}.

\end{enumerate}

The points 2 and 3 were discussed in the  literature, while the properties
of the GE interaction needs some comments and the behaviour of $V_{\rm
GE}$ at large distances, called the color Coulomb screening effect,
was studied first in Ref.~\cite{38} and recently in Ref.~\cite{27}.

In the simplest treatment \cite{38} a deformation of the confining
film, owing to propagating the quark and the antiquark trajectories,
was not optimal and due to the transformation of a gluon into a one-gluon
glue lump \cite{23}, the screening of the GE interaction, $V_{\rm OGE} = -
\frac{4\alpha_{\rm v}}{3 r} f(\rm scr)$, was shown to exist already
at  distances $\sim 0.6$~fm. Notice that such a strong screening
is not seen in bottomonium, where  $\chi_b(2P)$ and $\Upsilon(3S)$
have sizes $\sim 0.6$~fm and 0.7~fm \cite{25,39}.

In a more accurate treatment \cite{27} one has to maintain
full gauge invariance of the OGE interaction, and in addition
take into account the Wilson criterium of the minimal area law of
the resulting film surface, which contains both OGE and confinement.
Denoting the time distances between consequent gluon-exchanges as $L$,
at large $L$, one can consider this system as a hybrid excitation of the
$q\bar q$ system of size $L$.  Then, the mass
of a transverse excitation is $m_{\rm scr}\cong \frac{\sqrt{12}}{L}$
\cite{40}. On the other hand, the average value of $L$ enters into the
action (the total Lagrangian) exponent as $ \exp( - V_{\rm OGE}~L) \sim{\cal O}(1)$,
or $L^{-1}\approx \frac{4\alpha_V}{3 r}$.  As a result one obtains
an estimate of the screening mass (if $\alpha_{\rm V}\cong 0.50$)
\cite{27},
\be
 m_{\rm scr} \approx  \sqrt{12}\, \frac{4\alpha_V}{3 r_{\rm eff}}\la  0.40~{\rm GeV}
 \quad (r_{\rm eff}\ga 1~{\rm fm}).
\label{eq.44}
\ee
Note that this estimate refers to large distances, $r\ga 1$~fm,
where the deformation of the surface, due to  the gluon exchange, is
significant, while  the screening (and deformation) is suppressed
for the smaller $L$.

We may conclude that the flattening of the CP and the screening of
$V_{\rm GE}$ start at approximately the same distances, $r >
(1.0-1.2)$~fm, but their nature is different. The flattening effect
appears due to creation of light $q\bar q$ holes, which,
decreasing the film surface (or the Wilson loop), also decreases
the string tension at large distances.

The screening of the GE potential occurs because the movement of
the gluon is restricted inside the film surface and gives rise to a
deformation of the film, so that due to confinement a kind of
``gluon mass'', $m_{\rm scr.}\sim 0.4$~GeV,  appears.  Thus the
analysis of Ref.~\cite{27} does not support the idea that the
screening of the CP and the GE potential have the same origin and
the same screening function can be used for both potentials as
suggested in Ref.~\cite{33}. For that reason, here the GE potential
without screening is taken, while in previous studies the small
$\alpha_{\rm V}=0.30$ (a kind of screening) was used \cite{15} and
the GE potential with exponential screening function
$\exp(-\delta\, r)$ ($\delta=0.20$~GeV) was taken in Ref.~\cite{16}.

We also assume that the light mesons can be described by the universal
GE potential (\ref{eq.11}) with the same parameters as in heavy
quarkonia and heavy-light mesons \cite{25,39}, where the value of
the vector coupling $\alpha_{\rm V}(n_f)$ is determined by the QCD
vector constant $\Lambda_{\rm V}(n_f)$, which is defined through
the QCD constant $\Lambda_{\overline{MS}}(n_f)$ \cite{41}:
\be
 \Lambda_{\rm V}(n_f) = \Lambda_{\overline{MS}}(n_f)
 \exp \left(-\frac{a_1}{2\beta_0} \right),
\label{eq.45}
\ee
where $\beta_0=11- \frac{2}{3} n_f, ~a_1 = \frac{31}{3} - \frac{10}{9} n_f$. For $n_f=3$ it gives
\be
\Lambda_{\rm V} (n_f=3) = 1.4753\,\Lambda_{\overline{MS}}(n_f=3).
\label{eq.46}
\ee

In pQCD the QCD constant $\Lambda_{\overline{MS}}(n_f=3)=0.339(10)$~GeV
is now known from the analysis of $\alpha_s(n_f)$, where the coupling
$\alpha_s(M_Z)=0.1184(7)$ was taken as input and the matching
procedure at the $b$-quark mass ($n_f=5$) and the $c$-quark mass
($n_f=4$) was performed \cite{24}. Then the value
$\Lambda_{\overline{MS}}(n_f=3)=339(10)$~MeV is obtained, which is
significantly larger than that used in the past \cite{3}. With the use
of the relation (\ref{eq.46}) the value, $\lambda_{\rm V}(n_f=3)
=500(15)$~MeV, follows and this large number has a small
uncertainty. However,  since the value $\Lambda_{\overline{MS}}(n_f=3)$
depends on the $b$- an $c$-quark masses, taken at the matching
points, we expect that the uncertainty may be larger and this
statement is confirmed in the analysis of the bottomonium spectrum
\cite{25}, where a smaller $\Lambda_{\rm V}(n_f=3)=480(20)$~MeV was
shown to provide the best description of the bottomonium spectrum,
if the IR regulator $M_{\rm B}=1.15$~GeV is used. Here in our study of
the light meson spectra the preferable value of $\Lambda(n_f=3)$,
which does not contradict the description of the bottomonium spectrum,
is $\Lambda_{\rm V}(n_f=3)=460\pm 20$~MeV.

In coordinate space the strong vector coupling $\alpha_{\rm V}(r)~
(n_f=3)$ is expressed via the vector coupling $\alpha_{\rm V}(q^2)$
in the momentum space \cite{16,25},
\be
 \alpha_{\rm V}(r) = \frac{2}{\pi}\int \limits_0^\infty {\rm d} q
 \frac{\sin(qr)}{q} \alpha_{\rm V}(q^2),
\label{eq.47}
\ee
which is taken in the two-loop approximation,
\be
 \alpha_{\rm V}(q^2) = \frac{4\pi}{9 t} \left(1 - \frac{64}{81} \frac{\ln t}{t}\right),
\label{eq.48}
\ee
with  $t=\ln[ (q^2 +M_{\rm B}^2)/\Lambda_{\rm V}^2]$. The parameters
$\Lambda_{\rm V}$ and $M_{\rm B}$ are taken from Ref.~\cite{16},
\be
 \Lambda_{\rm V}(n_f=3)=0.465~{\rm GeV}, \quad M_{\rm B}=1.15~{\rm GeV},
 \quad \alpha({\rm asym})=\alpha_{\rm V}(r\rightarrow \infty)=0.571,
 \quad \frac{4}{3}\alpha_{\rm asym}=0.761.
\label{eq.49}
\ee
where for $\Lambda_{\rm V}(n_f=3)=0.465$~GeV the frozen (asymptotic)
coupling $\alpha_{\rm asym}=0.571$, or $e_{\rm asym}=0.761$, and
just this value was used in our analysis of the RTs in Sections \ref{sec.3} and \ref{sec.4}.
From Eq.~(\ref{eq.47}) it can be derived that the asymptotic values
of the $\alpha_{\rm V}(q^2)$ and $\alpha_{\rm V}(r)$ coincide and
the behaviour of $\alpha_{\rm V}(r)$ for two values of the constant, $\Lambda_{\rm
V}(n_f=3)= 0.465$~GeV and 0.50~GeV, respectively,  with $M_{\rm B}=1.15$~GeV, is
shown in Fig.1.

\vspace{2ex}
\begin{figure}
\begin{center}
\caption{\label{Fig.1} The GE  potential in $r$-space, for two values of $\Lambda_{\rm V}(n_f = 3)$:
solid curve $\Lambda_{\rm V}= 0.465$ GeV, dashed curve $\Lambda_{\rm V} = 0.500$ GeV. }
\vspace{2ex}
\includegraphics[width=60mm]{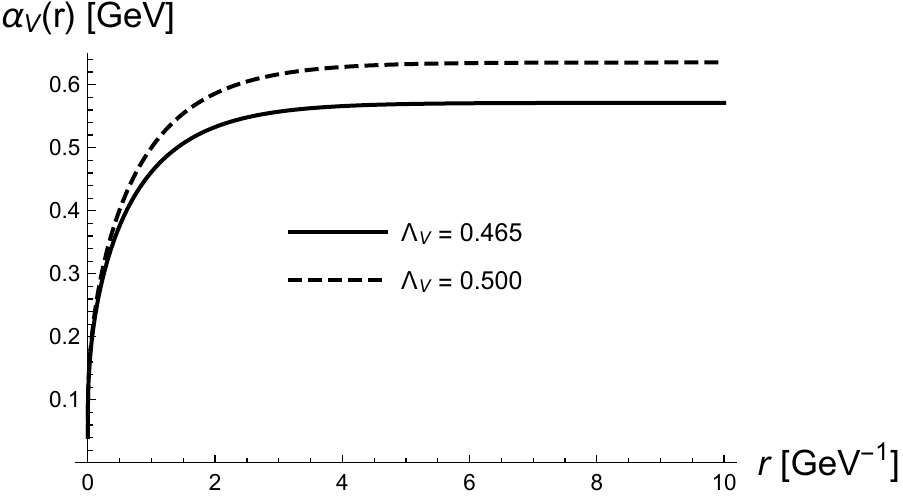}
\end{center}
\end{figure}
\vspace{2ex}

As seen from Fig.1, the AF effect is important only for the $1S$, $1P$, and $2S$
states, which sizes are $\la 1.1$~fm, and the following effective coupling, $\langle \alpha_{\rm V)}(r)\,
r^{-1}\rangle _{nl}= \alpha_{\rm eff} \langle r^{-1}\rangle _{nl}$  are defined:
\begin{eqnarray}
\alpha_{\rm eff.}(1S) & = & 0.48, ~e_0(1S)=0.64, ~ \alpha_{\rm eff.}(1P)=
 0.495, ~e_l(1P)=0.66,
\nonumber \\
\alpha_{\rm eff.}(2S) & = & 0.49~(n\geq 1),~ e_0(2S)=0.65, ~\alpha_{\rm eff.}=
 0.75(1) ~(\l\geq 2,~n\geq 1).
\label{eq.50}
\end{eqnarray}
As shown in Sect.~\ref{sec.3}, the use of the effective coupling allows to
present the physical picture in a clear way and for excited states provides the values
of $\delta_{\rm GE}(nl)$, given in Table~\ref{tab.03}, with accuracy
$\sim (10-15)$~MeV.

\section{The flattened potential $V_{\rm f}(r)$}
\label{sec.6}

In the flattened CP $V_{\rm f}(r)$ the string tension depends on $r$,
\be
V_{\rm f}(r)= \sigma_{\rm f}(r) r, ~\sigma_{\rm f}(r)= \sigma (1 -  \gamma f(r)),
\label{eq.51}
\ee
where  $\sigma(r)$ is defined by three parameters: first, the
characteristic distance $R_0\sim (1.2-1.4)$~fm,  where the flattening
effect starts and string breaking becomes possible; its value is
taken from the lattice calculations \cite{22}. The second  parameter,
$\gamma$, determines the derivative of $V_{\rm f}(r)$ and the
asymptotic value of the string tension,
\be
 \sigma_{\rm f}(r\rightarrow \infty) = \sigma ( 1 - \gamma ).
\label{eq.52}
\ee
The variation of the parameter $\gamma$ in the range (0.30-0.50)
has confirmed the result of Ref.~\cite{16} that $\gamma=(0.40-0.45)$
provides the best description of the spectrum (see Fig. 2, where the
flattened CP is shown for $\gamma=0.40$).

\begin{figure}[htb]
\begin{center}
\caption{\label{Fig.2} The confining potential $V_{\rm f}(r)$  in GeV in $r$-space, $r$ in GeV$^{-1}$
is shown with  the
parameters from Eq.~(\ref{eq.54}), solid line,  and the potential $V_{\rm scr}(r)$ of Li and Chao, defined by
Eq.~(\ref{eq.55}), dashed line.}
\vspace{2ex}
\includegraphics[width=100mm]{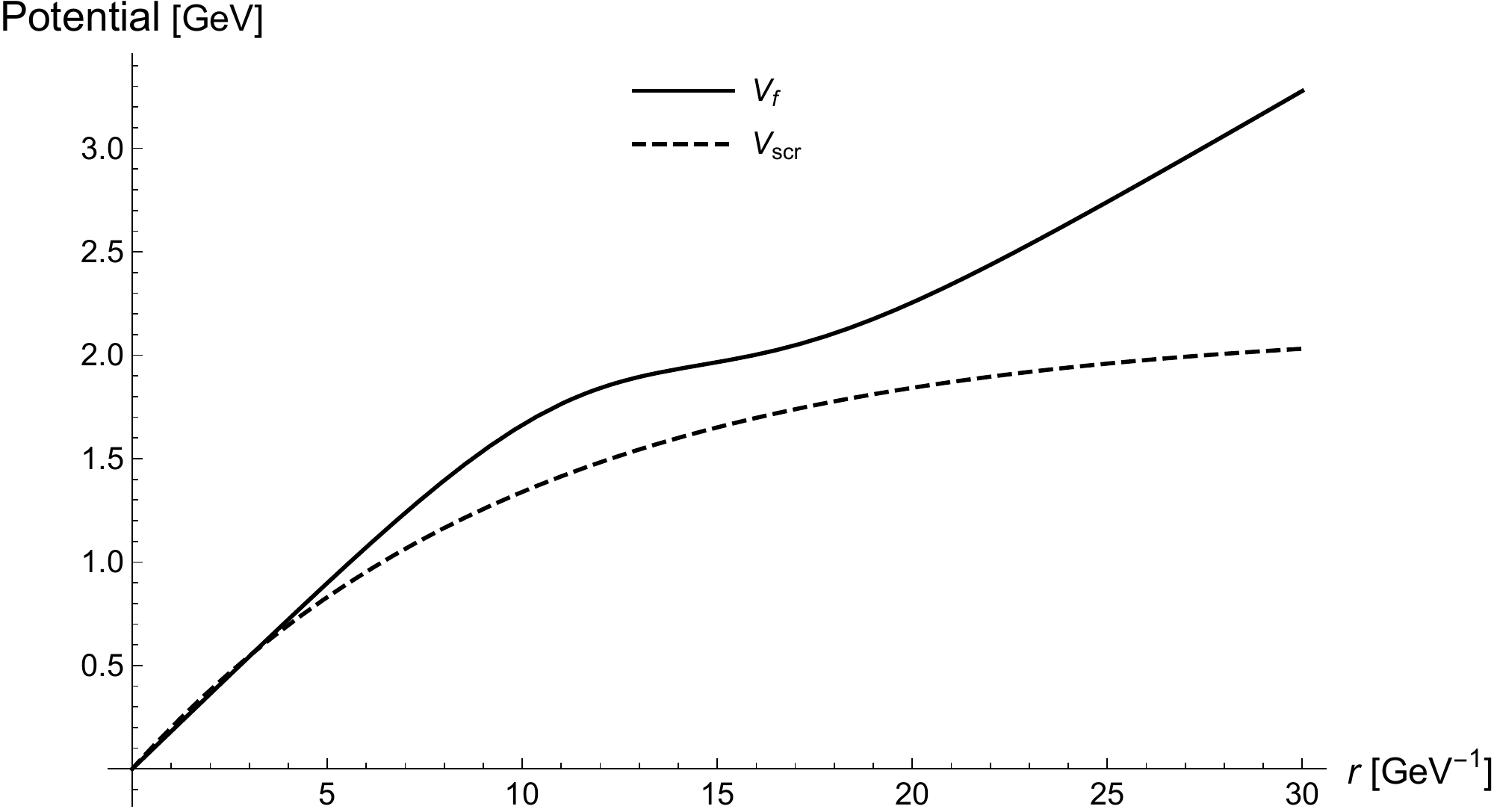}
\end{center}
\end{figure}

The third parameter, the
constant  $B$, enters the function $f(r)$,
\be
 f(r) = \frac{\exp(\sqrt{\sigma}(r -R_0))}{ B + \exp(\sqrt{\sigma}(r-R_0))}.
\label{eq.53}
\ee
where $f(r=R_0)=(B+1)^{-1}$ at the point $r=R_0$ shows how fast the
increasing function $f(r)$ is approaching  its asymptotic value,
$f_{\rm asym}=1.0$, at large distances, $r\sim 3.0$~fm. In other
aspects $f(r)$ can be rather arbitrary.

In our analysis all parameters were varied in wide ranges:
$\gamma=0.35-0.50$,~ $B=15-25$,~$R_0=(5 - 8)$\,GeV$^{-1}=(1.0 - 1.6)$\,fm,
for which   $\sigma_{\rm f}({\rm asym})=(0.090-0.11)$~GeV$^2$.
The best description of the spectrum is reached for the parameter values
\be
 \gamma=0.40, \quad B=20 , \quad R_0=6.0\,{\rm GeV}^{-1},
 \quad \sigma=0.182\,{\rm GeV}^2.
\label{eq.54}
\ee
From the physical point of view it is important that  at large
distances the chosen  potential, $V_{\rm f}(r) \to \sigma_{\rm
f\, asym}\, r$ becomes again linear with small $\sigma_{\rm
f\, asym}$ and therefore provides a quark and anti-quark to be
confined in a meson. This property of the flattened CP $V_{\rm
f}(r)$ differs from that of a screened CP $V_{\rm scr}(r)$, used
in many papers \cite{33,34,42},
\be
 V_{\rm scr}(r) = \lambda~r F_{\rm scr}, ~~F_{\rm scr}=
 \frac{1-\exp(-\delta r)}{\delta r},~
 ~\lambda=0.21~{\rm GeV}^2, ~\delta=0.0979~{\rm GeV}
\label{eq.55}
\ee
which has a linear behavior with large string tension $\lambda=0.21$~GeV$^2$
at $r < 0.5$~fm and then the screening effect starts already at
$r\sim 0.5$~fm.  At large distances   $V_{\rm scr}$ is approaching a constant,
$V_{\rm scr}(\rm asym.) =
\frac{\lambda}{\delta}=2.145$~GeV and therefore the quark
and antiquark inside a meson are not confined. The behaviour of
$V_{\rm scr}(r)$ is compared with $V_{\rm f}(r)$ in Fig. 2.  Notice
that the screening function $F_{\rm scr}$ cannot be used in the GE
potential, since it produces the interaction $V_{\rm GE}({\rm scr})=
- \frac{e}{\delta}\, \frac{1 -\exp(-\delta\,r)}{r^2}$, in which the main
term, proportional to $r^{-2}$ with very large Coulomb constant
$\frac{e}{\delta}\sim 5.0$, produces an unstable (unlimited) spectrum
and cannot be used.

In Eq.~(\ref{eq.54}) the string tension $\sigma=0.182$~GeV$^2$ is chosen
to keep for the $1S$ state the averaged string tension, $\langle
\sigma_{\rm f}(1S) \rangle = 0.180$\,{\rm GeV}$^2$, as in the purely linear
CP. From our point of view the phenomenological potential $V_{\rm
f}$ represents the physical picture rather well, but it has a
negative feature, namely, the non-monotonic behavior near the point
$r=R_0$, and to obtain a smooth behavior of some matrix elements,
a numerical regularization is needed.

With the flattened potential the spectrum significantly changes,
as compared to the linear CP. First, the sizes $ \sqrt{\langle
r^2(nl) \rangle}$ of the excited states increase and can reach  $\sim
2$~fm even for the $2D$ and $3S$ states (see Table \ref{tab.05}), although
the sizes of the low states ($1S,1P$) remain not large, in particular,
the r.m.s of the $\rho(1S)$ meson, $\sqrt{\langle r^2(1S)\rangle}=0.71$~fm
(if the GE potential is present), is in good agreement with predictions
in other approaches \cite{43} and for these states the linear CP
can be used. Secondly, due to the  flattening effect  the e.v.s
$M_{0{\rm f}}(nl)$ can be approximated like that in Eq.~(\ref{eq.21}) and
for the set of parameters Eq.~(\ref{eq.54}) and $\gamma=0.35,0.40,0.45$
the squared masses $M_{0{\rm f}}^2(nl)$ with $n\geq 1$ can be presented
as,
\begin{eqnarray}
 M_{0{\rm f}}^2(nl) ({\rm in~GeV}^2) & =  & 0.86(5)\, l  + 1.24(10)\, n  +2.51(10), \quad (\gamma=0.45),
 \nonumber\\
 M_{0{\rm f}}^2(nl)({\rm in~GeV}^2) & =  & 0.95(5) \, l  + 1.25(13)\, n + 2.64(10), \quad (\gamma=0.40,)
 \nonumber\\
 M_{0{\rm f}}^2(nl) ({\rm in~GeV}^2) & = & 1.04 (8) \, l  +  1.35(8) \, n  + 2.65(6), \quad (\gamma=0.35).
\label{eq.56}
\end{eqnarray}
where in all cases $n\geq 1$ and the ground states ($n=0$) are not
included, since their masses are determined by the linear CP.  In
Eq.~(\ref{eq.56}) one can see that the orbital and radial slopes
differ by $\sim 35\%$ for three different values of $\gamma$, and  the orbital slope $\beta_l$
and the radial slope $\beta_n$ increase for the smaller value of $\gamma$; also the
values of both slopes are significantly smaller than $\beta_l=1.44$~GeV$^2$ and
$\beta_n=1.70$~GeV$^2$ in the linear potential. This effect mostly occurs, because
in the higher excitations the averaged  $\langle \sigma(nl) \rangle$ is  $\sim (20-30)\%$
smaller than the string tension $\sigma=0.18$~GeV$^2$. It is of interest to notice
that in the case with $\gamma=0.40$  calculated here $\beta_l$ and $\beta_n$
turn out to be very close to the values obtained in the analysis
of the experimental data in Refs.~\cite{11,12}.

An unexpected result refers to the intercept, which in Eq.~(\ref{eq.56})
is very large, $\cong 2.6$~GeV$^2$, being even larger than
in the linear CP. It means that in the flattened CP the
GE and the SE corrections remain very important, while the string corrections
is rather small. These corrections are
defined by the same general formulas Eqs.~(\ref{eq.13},\ref{eq.14}),
if  there the kinetic energy $\mu(nl)$ and $\sigma$  are  replaced
by the m.e.s $\langle \mu_{\rm f} \rangle$ and $\langle \sigma_{\rm
f} \rangle$, respectively. However, the relations (\ref{eq.19}) and
(\ref{eq.20}) for the m.e.s $\langle r^{-1} \rangle$  and $\mu_{\rm f} (nl)$ are
not valid anymore and they have to be calculated in every case separately. In
Table \ref{tab.05} we compare the r.m.s.  in the linear CP with that in the
flattened CP (FCP) and  FCP+GE potential, and  show  in the
FCP, with or without the GE term, the sizes of the states with
$n\geq 2$ strongly increase.

\begin{table}[!htb]
\caption{The r.m.s.(in fm) and $\langle r^{-1}(nl) \rangle $ (in GeV) of light mesons
($m_q=0$) for the  linear potential (LP) ($\sigma=0.18$~GeV$^2$)
and the flattened confining potential (FCP) with the parameters
Eq.~(\ref{eq.54}), and for  the FCP + GE potential with the parameters
from Eq.~(\ref{eq.49}) \label{tab.05}}

\begin{center}
\begin{tabular}{|c|c|c|c|c|}\hline
     state &   r.m.s (LP) & r.m.s. FCP & r.m.s. FCP+GE &  $\langle r^{-1} \rangle$~(FCP) \\
\hline
     $1S$    & 0.82      &   0.86    &    0.71    & 0.357\\
     $2S$    &  1.47     &   1.53     & 1.30   & 0.288\\
     $3S$    & 1.65       &   2.42   &  2.12   & 0.204  \\
     $4S$    &  1.78      &   2.67   &  2.61   &  0.193 \\
     $5S$    &  2.08      &   2.94    & 2.79 & 0.189 \\
\hline
     $1P$    &  1.06      &   1.13    & 1.00    & 0.226 \\
    $2P$     &  1.43      &   1.95     & 1.69  & 0.176 \\
    $3P$     &  1.72     &    2.64    &  2.53  & 0.147\\
    $4P$     &  1.97     &   2.78     &  2.69  & 0.146 \\
\hline
    $1D$     &  1.24       &1.41     &   1.28  & 0.172\\
    $2D$     &  1.56    & 2.42    &   2.18    & 0.134\\
    $3D$     &  1.83     & 2.70     &   2.67  & 0.130\\
    $4D$     &  2.06     & 2.94     &  2.83  &  0.127\\
\hline
    $1F$   & 1.41       & 1.77   &  1.59  & 0.134\\
    $2F$    & 1.70       &2.73   &   2.61  & 0.115\\
\hline
\end{tabular}

\end{center}
\end{table}
An interesting feature of $V_{\rm f}(r)$ refers to the averaged
m.e.s $\langle r^{-1}(nl) \rangle$ (see Table~\ref{tab.05}) and
to $\langle \sigma_{\rm f}(nl) \rangle$~ (see Table
\ref{tab.06}), which for excitations with $n > 1$ coincide within
$5\%$ accuracy (if $l \geq 2$). Also, in the flattened CP the kinetic
energies $\mu_{\rm f}(nl)$, as a function of $n$, grow slowly
and have about $100-200$~MeV  smaller values than $\mu_0(nl)$
in the linear CP. Due to this feature the self-energy
correction, proportional to $\mu_{\rm f}^{-1}$, remains large, about
$ -(240-300)$~MeV, and very important for high excitations. Also, in the presence
of the  GE potential the kinetic energy increases slowly,  by $ \sim (5-10)\%$,
(see Table \ref{tab.07}) and in some cases the difference between them
can be neglected.

\begin{table}[!htb]
\caption{The averaged $\langle \sigma_{\rm f} \rangle_{\rm nl}$ (in GeV$^2$) for
the flattened CP with the parameters Eq.~(\ref{eq.54})
\label{tab.06}}
\begin{center}
\begin{tabular}{|c|c|c|c|c|}
\hline
 $ n/l$   &  0      &  1      &   2    &   3  \\
 \hline
   1      &  0.173  & 0.167  & 0.155 & 0.150  \\
   2      & 0.162  & 0.165  & 0.150 & 0.148  \\
   3      & 0.150   &0.158  & 0.148 & 0.146  \\
\hline
\end{tabular}
\end{center}
\end{table}

\begin{table}[!htb]
\caption {The kinetic energies $\mu_{\rm g}(nl)$ (in MeV) for the potential
$V_{0{\rm f}}(r)=V_{\rm f}(r) +V_{\rm GE}(r)$
\label{tab.07}}
\begin{center}
\begin{tabular}{|c|c|c|c|c| }\hline
  $n/l $ &  0     &  1      & 2    &     3  \\
\hline
  0    & 400   & 491   & 539     & 564   \\

  1    & 460   & 480    &  525    & 580  \\

   2   & 520  &  482   &   536 & 594   \\

  3    &  550  &  500  & 560  &  620  \\
\hline
\end{tabular}
\end{center}
\end{table}

In the flattened CP the m.e.s $\langle r^{-1}\rangle_{nl}$ are small and practically equal
(for high excitations), and therefore they cannot be expressed via the factors $A(nl)$
(\ref{eq.20}) and
$Z(nl)$ (\ref{eq.25}). Moreover, these m.e.s are not proportional to the e.v.s $M_{0{\rm f}}(nl)$
and this fact changes the physical picture. To calculate the GE correction  the general form
$\delta_{\rm GE}(nl)= - e_{\rm eff}\langle r^{-1} \rangle_{nl}$ has to be used, and in the string
and the SE corrections the averaged string tensions, which are different in the states with
different $l$~(and fixed $n$),
have to be taken. Since the GE correction is small, the orbital and the radial slopes of the
$M_{\rm g}^2(nl)$-trajectory, where $M_{\rm g}=M_{0{\rm f}}+\delta_{\rm GE}$, practically
do not  change (see their values in Eq.~ (\ref{eq.56}) for $\gamma=0.40$):
\be
 M_{\rm g}^2 (nl)~({\rm in~ GeV}^2) =  0.94(4)\, l  + 1.24(9)\, n + 2.17(7).
  ~(l\not= 0,~\gamma=0.40,~n\geq 1).
\label{eq.57}
\ee
Thus, in the flattened CP + GE potential the parameters of the $M_{\rm g}^2 (nl)$-trajectory 
appear to be only $(1-3)\%$ smaller than those in the purely flattened CP.

In Tables~ \ref{tab.08} and \ref{tab.09} besides the GE corrections, we give also the e.v.s
$M_0(nl)$
of Eq.~(\ref{eq.7}) with the linear CP and the mass shifts, produced by the flattening effect:
$\Delta_{\rm f}(nl)=M_0(nl) -M_{0{\rm f}}(nl)$, which are large, $\sim - (300-350)$\,MeV.
As seen from  Eq.~(\ref{eq.57}), the masses $M_{\rm g}(nl)$, defined without the self-energy
correction, are still  $\sim (300-400)$\,MeV larger than their
experimental values in  Table~\ref{tab.09} and
we come to the conclusion that the use of the flattened CP and the GE correction cannot
provide the correct intercept, because the self-energy correction plays a dominant
role to decrease its value.

In Table \ref{tab.08} the masses $M(n\,^3S_1)$  of the $n\,^3S_1$ states (for the
set of the parameters (\ref{eq.54})) together with the mass shifts
due to the flattening effect,  $\Delta_{\rm f}(nl)=M_{0{\rm f}}(nl)
- M_0(nl)$, and all corrections, including the hyperfine correction
$\delta_{\rm hf}$, are given.
\begin{table}[!htb]
\caption{The eigenvalues  $M_0(nS)$ (in MeV) of  Eq.~(\ref{eq.7})
with the linear potential, the shifts $\Delta_f(nS)= M_{0{\rm f}}(nS)-M_0(nS)$,
the corrections $\delta_{\rm SE},~\delta_{\rm GE},~\delta_{\rm hf}$,
and the masses $M(n\,^3S_1)$ \label{tab.08}}
 \begin{center}
\begin{tabular}{|c|c|c|c|c|c|c|c| }
\hline
State & $M_0(nS)$ & $\Delta_{\rm f}(nS)$& $\delta_{\rm SE}(nS)$ & $\delta_{\rm GE}(nS)$ &
 $\delta_{\rm hf}(nS)$ & $M(n^3S_1)$& $M({\rm exp.}) [25]$\\
\hline
$1\,^3S_1$ & 1339 & 0    & -405 & -225 & 66 &  775 & $\rho(775),~M=775.5(3) $\\

$2\,^3S_1$ & 1998 & -55 & -338  & -190 & 40 & 1455 & $\rho(1465), M=1465(25)$\\

$3\,^3S_1$ & 2498 & -198 & -291 & -143 & 26 & 1892 & $\rho(1900),~M=1880(30)$\\

$4\,^3S_1$ & 2915 & -346 & -244 & -131 & 20 & 2214 & $\rho(2150),~M=2254(22)$\\
\hline
\end{tabular}
\end{center}
\end{table}
The masses of the $n\,^3S_1$ states agree
with the experimental values, with exception of the mass of $\rho(4S)$,
which value is not well established yet \cite{10}; notice, that the
calculated mass,  $M(\rho(4S))=2214$~MeV, is in agreement with the
BaBar data \cite{44}. Taking the masses from Table~\ref{tab.08}, the slope
$\beta_n(l=0)$ of the $\rho(n\,^3S_1)$ trajectory,
\be
 \beta_n(l=0) = 1.43(11)~{\rm GeV}^2,
\label{eq.58}
\ee
is obtained, which because of a large uncertainty can be considered to be approximately linear.
Nevertheless, this radial slope is in good agreement with the experimental
$\beta_n(l=0,{\rm exp.})= 1.47(13)$, if the following experimental
masses: $M(\rho(1S)=775$~MeV,~ $\rho(2S)=1.465(25)$~MeV,~
$\rho(3S)=1890(20)$~MeV, and  $M(\rho(4S)=2240$~MeV \cite{10}, are used.

The masses of the orbital excitations $M_{\rm g}(nl)$ are given in Table~\ref{tab.09},
where one can see that in the high excitations  ($l\not=0,~n\geq 2$) the
shifts due to flattening, $\Delta_{\rm f} \sim -300$~MeV, are very
large, while the GE corrections, $\sim -90$~MeV, are relatively
small. However, without the SE corrections, the masses,
$M_{\rm g}(nl) = M_0(nl) + \Delta_{\rm f}(nl)+ \delta_{\rm GE}(nl)$,
($l\geq 1,~n\geq 1$) exceed by $\sim (300-400)$~MeV  the experimental values.

\begin{table}[!htb]
\caption{The eigenvalues $M_0(nl)$ of Eq.~(\ref{eq.5}) with linear CP,
the mass shifts $\Delta_f (nl)$, the GE corrections $\delta_{\rm GE}(nl)$
(in MeV) and $M_{\rm g}(n,l)=M_0+\Delta_{\rm f} + \delta_{\rm GE}$ \label{tab.09}}
 \begin{center}
 \begin{tabular}{|c|c|r|r|c|l|c}
 \hline
 state & $M_0(nl)$ & $\Delta_{\rm f}(nl)$ & $\delta_{\rm GE}(nl)$ & $M_{\rm g}(nl)$ & Exp. \cite{10}\\
 \hline
$1P$ & 1792 &  0 ~     & -140 ~  & 1652  & $a_2(1320)$\\

$2P$ & 2315   & -102~   & -144~ & 2069   & $a_2(1700)$\\

$3P$ &  2750   & -278~  &  -123~ &  2349  & absent \\

$4P$ & 3129  &  -398~   & -114~  &  2617 & absent \\
\hline
$1D$ &  2155  & 0~  &  -124~ & 2031   & $\rho_3(1700)$\\

$2D$  &  2601  & - 173~  &  -113~ & 2315 & $ \rho(1990)$\\

$3D$  & 2990  & -342~  & -96~   &  2552  & absent \\

$4D$ & 3337&  -448~   & -94~  & 2795   &  absent \\
\hline
$1F$ &2465  & 0~    & -97~  &   2368   & $a_4(2040)$\\

$2F$ & 2861& -256~    &-86~ &   2519  & absent \\

$3F$ & 3215   &-394~ &  -84~&  2737 & absent  \\
\hline
\end{tabular}
\end{center}
\end{table}

\section{The universal Regge Trajectories}
\label{sec.7}
In the previous section it was shown that in the flattened potential plus the GE correction,
the masses $M_{\rm g}(nl)$ are larger than the experimental
values by (300-400)~MeV, and other corrections have to be taken into account. The string corrections
$\delta_{\rm str}$, defined by the Eq.~(\ref{eq.14}), depend  on the m.e.s
$\langle r^{-1} \rangle_{nl}$, while the expressions, Eqs.~(\ref{eq.19}) and (\ref{eq.20}), are not
valid anymore and here the exact values of $\langle  r^{-1} \rangle_{nl}$ from Table ~\ref{tab.05}
(and  the kinetic energy $\mu_{\rm g}(nl)$ from Table~ \ref{tab.07}) are used. The values of the string correction,
\be
 \delta_{\rm str}(nl)= - l (l+1)\langle \sigma(nl) \rangle
  \frac{ \langle r^{-1} \rangle_{nl}}{8 \mu_{\rm g}^2(nl)}.
\label{eq.59}
\ee
are given in Table~\ref{tab.10}. In high excitations they are small, $\sim -(30-50)$~MeV.
On the contrary, the SE corrections remain large even in high excitations,
where also the centroid masses $M_{\rm cog}(nl)$ are given. From this table one can
also see that the centroid masses agree with the experimental masses,
although the fine-structure splittings were not taken into account.
\begin{table}[!htb]
\caption{The centroid masses $M_{\rm cog}(nl)=M_{\rm g}(nl) +
\delta_{\rm SE}+\delta_{\rm str}$ (in MeV), calculated with the
parameters, Eq.~(\ref{eq.49}),  and  $\gamma=0.40 $ \label{tab.10}}
\begin{center}
\begin{tabular}{|c|c|c|c|c|c|c|}
\hline
state     & $M_{\rm g}(nl)$   &  $\mu_{\rm g}(nl)$ & $\delta_{\rm SE}$ & $\delta_{\rm str}$&  $M_{\rm cog}(nl)$ &    Exp.  [10]
\\
\hline
$1P$  & 1652       & 491       & - 330      & -40     & 1282    &  $a_2(1320)$\\
$2P$  & 2069       & 480       & -310       & -33     & 1726    & $a_2(1700)$ \\
$3P$  & 2349       & 482       & -291       & - 30    & 2028    &  absent \\
$4P$  & 2617       & 500       & -268       & - 25    & 2324    & absent  \\
\hline
$1D$   & 2031      & 539       & -292       & -77     & 1662    & $\rho_3(1690)$ \\
$2D$   & 2315      & 525       & -262       & -60     &  1993   & $\rho_3(1990)$\\
$3D$   & 2552      &536        & -257       & - 51    & 2244    &  absent \\
\hline
$1F$   & 2368      &564         & -270       & - 107  & 1991   & $a_4(2040)$  \\
$2F$   & 2519     & 580         & -231       & -80     &  2208  &  absent \\
$3F$   & 2737     &  594        & -221       & - 74    & 2442   &  absent   \\
\hline
\end{tabular}
\end{center}
\end{table}

Then using the squared masses, the RT trajectory in the ($M_{\rm cog}^2,nl$) plane
($l\not= 0$) can be built,

\be
 M_{\rm cog}^2(n,l) ({\rm in~GeV}{}^2) = 1.03(9)\, l+  1.15(9) \, n + 0.65(15),
\label{eq.60}
\ee
where the orbital and the radial slopes have rather close values and even coincide
within the theoretical errors. However, the central value of the orbital slope of this RT
(\ref{eq.60}) is $\sim 10\%$ smaller than that of the leading RT (\ref{eq.43}) and
this difference illustrates the accuracy of our calculations with the flattened potential.
In Table~\ref{tab.11} we compare the masses of the high excitations with $l\not= 0$, described
by the RT (\ref{eq.60}), and those given in Table~\ref{tab.10}. One can see that in most cases
the agreement is better than 30~MeV.

We have chosen here the phenomenological flattened CP (\ref{eq.51}), but one cannot exclude
that the true (``ideal") flattened CP is different, in particular, because in our case the matching
of the linear
and the flattened CP is not smooth. Therefore, equal values of the radial and
orbital slopes are not excluded either. However, in our analysis the calculated RT can be
called approximately universal.

It is important to stress that in the flattened CP plus the GE potential with the strong
coupling, $\alpha_{\rm asym} =0.57$, the masses of the high excitations are too large and only
due to the self-energy correction  correct values of the masses are obtained.
Our analysis also has shown that in the flattened CP the role of the GE interaction is less important
and therefore one cannot draw a definite conclusion whether at large distances a strong
screening of GE potential exists, or not. This statement is supported
by our result that the RT in purely flattened CP with
$\gamma=0.40$ (\ref{eq.56}) and the RT (\ref{eq.57}), where in the masses
the GE correction is taken into account, have  practically equal
$\beta_l =0.95(5)$~GeV$^2$ and radial slope, $\beta_n\sim  1.25(14)$~GeV$^2$.

To draw the conclusion whether the GE potential is screened or not, it is more
perspicuous to study not very high excitations of light mesons, but to
concentrate at lower resonances with $l=0,1$,  which  sizes are not very large,
$\sim 1$~fm, and where the GE correction is more important. Also the information about
the GE interaction at large distances can be extracted from the study of high excitations
(with $n\geq 2$) in charmonium, or  the bottomonium resonances above the $B\bar B$ threshold.
Notice that in charmonium the flattening effect is smaller than in light mesons \cite{39},
but the GE correction is larger.

\begin{table}[!htb]
\caption{Comparison of the masses $M_{\rm cog}$ (in MeV) from Table
~\ref{tab.10} and those defined in the RT given in Eq.~(\ref{eq.60})
\label{tab.11}}
\begin{center}
\begin{tabular}{|c|c|c|c|  }
\hline
  State       &      $M_{\rm cog}(nl)$  & $M(nl)$ Eq.~(\ref{eq.60})&  Exp.\\
  \hline
   $1P$   &    1282    &    1315  &  $a_2(1320)$   \\
   $2P$   &     1726   &  1697  & $a_2(1700)$ \\
   $3P$  &      2028    &  2007  & absent   \\
   $4P$   &   2324       &  2276  & absent     \\
\hline
   $1D$    &    1662    &  1661  &   $\rho_3(1690)$ \\
   $2D$   & 1993  &   1977  &     $\rho_3(1990)$\\
   $3D$   & 2244  &  2249  & absent\\
\hline
   $1F$  & 1991  &  1977 &  $ a_4(2000)$\\
   $2F$  & 2244   & 2249  &  absent \\
   $3F$& 2442  & 2467 &  absent \\
\hline
\end{tabular}
\end{center}
\end{table}

\section{Conclusions}
\label{sec.8}
The spectrum of light mesons was studied with the use of the RSH
with the flattened confining potential (FCP), taking into account the
gluon-exchange (GE),  the self-energy (SE), and the string corrections. We have confirmed
that the flattening effect, existing due to the creation of light $q\bar q$-pairs,
produces large mass shifts, which can reach $\sim -300$\,MeV
for the $3P$ and $3D$ excitations, and the best set of the flattened
potential parameters was determined.  Our calculations show that
agreement with the experimental values of the masses can be reached,
if all corrections are taken into account, but only the self-energy
correction provides the linearity of the RT.

A special accent was placed on the role of the GE potential by
performing calculations with the universal GE potential without
screening, like the one  that is used in heavy quarkonia. Our
analysis has shown that for a weak GE potential (with strong
screening) it is not possible to describe the leading RT($n=0$),
while the masses of high excitations weakly depend on the GE corrections.
If the strong universal GE potential, as in heavy quarkonia, is taken,
then the light meson masses with $l \not= 0$ are described by the RT, where
the values of the orbital slope, $\beta_l=1.03(9)$\,GeV$^2$ and the radial slope,
$\beta_n=1.15(12)$\,GeV$^2$ are close,  thus this RT can be considered
as approximately universal and the predicted masses agree with the existing
experimental data.

However, in any case the $n\,^3S_1$ -trajectory does not
belong to this RT, since their masses are strongly affected by the
GE interaction  and the spin-spin interaction, providing a large
slope of the radial $\rho(n\,^3S_1)$ trajectory, $\beta_n\approx
1.43$~GeV$^2$.

\end{document}